\documentstyle[
12pt,
prd,epsf,aps,amsmath,amssymb,graphicx,
preprint
]{revtex}
\draft
\begin{document}
\title{Asymptotic power-law tails of massive scalar fields
in  Reissner-Nordstr\"{o}m background}
%\title{late-time behaviors of a massive scalar field 
%in Reissner-Nordstr\"{o}m background}
%\\
%version Aug.30, 2000}
\author{Hiroko Koyama\thanks{Email: hiroko@allegro.phys.nagoya-u.ac.jp}
 and Akira Tomimatsu\thanks{Email: atomi@allegro.phys.nagoya-u.ac.jp}}
\address{Department of Physics, Nagoya University, Nagoya 464-8602, Japan}
%\date{30 November, 2000}
\maketitle  
%%%%%%%%%%%%%%%%%%%%%  abstract  %%%%%%%%%%%%%%%%%%%%%%%%%%%%%%%%%%%%%%%%%%%%
 \begin{abstract}
We investigate dominant late-time tail behaviors
 of massive scalar fields in nearly extreme Reissner-Nordstr\"{o}m background.
It is shown  that 
the oscillatory tail of the  scalar fields 
has the 
decay rate of  $t^{-5/6}$
at asymptotically late times. 
The physical mechanism by which the asymptotic $t^{-5/6}$ tail
is generated
and 
the relation between the field mass and 
the time scale when the tail begins to dominate,
are discussed in terms of resonance backscattering due to spacetime curvature.
\end{abstract}
%, 04.62.+v, 04.70.Dy}
%%%%%%%%%%%%%%%%%%  Introduction  %%%%%%%%%%%%%%%%%%%%%%%%%%%%%%%%%%%%%%%%%%%
\section{Introduction}

%%BH$B$H%9%+%i!<>l$NAj8_:nMQ$rD4$Y$k$3$H$O(BBH$B$N@-<A$rCN$k$&$($G=EMW$G$"$k!#(B
%BH$BGX7J;~6u$N%F%9%H%9%+%i!<>l$N$U$k$^$$$+$i$"$kDxEYJ,$+$k!#(B
%$B0JA0$+$iB?$/(B
%$B$N?M$K$h$C$F8&5f$5$l$F$-$?!#(B
%$B$=$NKX$s$I$O!"(Bmassless $B$N>l9g$G$"$C$?!#(B
%%$B%9%+%i!<>l$,(Bmassless $B$N>l9g$K$I$N$h$&$J8=>]$,5/$-$k$+$K$D$$$F$O!"(B

Various interactions of black holes with scalar fields have been
extensively studied for a long time.
Though previous works have been mainly concerned 
with the evolution of a massless scalar field, 
the analysis of massive ones will be also physically important.
For example, in higher-dimensional theories, the Fourier modes 
of a massless scalar field 
behave like massive fields known as Kaluza-Klein modes,
and the recent development of the Kaluza-Klein idea 
(e.g., the Randall-Sundrum model \cite{R-Smodel} in string theory)
strongly motivates us to understand the evolutional features
due to the field mass in detail.

%Various interactions of black holes with scalar fields 
%have been extensively studied for a long time.
%Many of them are about massless scalar fields rather than massive fields.
%It has been found that
%In particular, it has been found that 
%massive scalar fields can cause phenomena
%qualitatively different from massless fields.
%The evolution of a massive scalar field is 
%important in Kaluza-Klein theories, in particular, 
%where the Fourier modes of a massless scalar field behave like 
%massive fields, the so-called Kaluza-Klein modes.
%The importance of Kaluza-Klein theories
%has been encouraged
%motivated 
%by
%the proposal of Randall-Sundrum model in string theory \cite{R-Smodel}
%these days.
% has been regarded as important
%by the proposal of Rendall-Sundrum model in string theory\cite{R-Smodel}.

Massive scalar fields in black hole spacetimes can cause 
interesting phenomena which are qualitatively different from massless case.
The remarkable example is the vacuum polarization
$\langle \phi ^2 \rangle$
of a quantum massive
scalar field $\phi$, which is in thermal equilibrium
with a 
nearly extreme Reissner-Nordstr\"{o}m black hole \cite{TandK}.
Because the black-hole temperature is very low, the mass-induced excitation of
$\langle \phi ^2 \rangle$ results in the resonant 
amplification at $mM \simeq O(1)$ for the field mass $m$
and the black-hole mass $M$.

Such a resonance behavior due to the mass of a field interacting
with a black hole may appear in various processes as a basic feature of
black-hole geometry.
For a step to support this conjecture, in this paper, we turn our attention
to a problem on time evolution of classical fields.

The evolution of a massive scalar field
in Schwarzschild background
was first analyzed by Starobinskii and Novikov, 
using the complex plane approach \cite{S-N}, and 
they found that because of the mass term,
there are poles in the complex plane closer to the real axis than 
in the massless case,
which lead to inverse power law behavior 
with smaller indices than massless case.
%Burko studied numerically the late time behavior of massive scalar fields in
%Schwarzschild for two cases:
%(a) test field over a fixed Schwarzschild background, and 
%(b) self-gravitating, fully-nonlinear evolution of a massive scalar field
%in spherical symmetry.
%In both cases he found that the evolution should be  derided to an 
%intermediate regime and true late time regime, 
%and oscillatory inverse power-law behavior.
Recently, 
%Hod and Piran \cite{HandP} 
it was pointed out that the late-time tails 
of massive scalar fields 
in Reissner-Nordstr\"{o}m spacetime
are quite different from massless fields
in the existence of the intermediate late time tails \cite{HandP}. 
If the field mass $m$ is small,
%In the case of small mass, 
namely $mM \ll 1$, 
the oscillatory inverse power-law behavior
\begin{equation}
  \label{eq:HP}
  \psi \sim t^{\scriptscriptstyle - l-\frac 32}\sin(mt),
\end{equation}
dominates as the intermediate late-time tails.
% \cite{HandP}.
Note that massless fields decay more rapidly without any oscillation,
as was first studied by Price \cite{Price}.
The analytical approximation (\ref{eq:HP}) was
 derived from the flat-spacetime approximation,
where the effects of spacetime curvature are neglected.
Though the behavior (\ref{eq:HP}) 
% analytical approximation (\ref{eq:HP}) 
was  numerically verified at
intermediate late times,
%On the other hand,
Hod and Piran \cite{HandP} also
%Hod and Piran \cite{HandP} also 
 suggested 
that another wave pattern dominates at very late times,
namely the intermediate tails (\ref{eq:HP}) are
not a final asymptotic behaviors,
and mentioned that  ``fields decay at late times
slower than any power law''.
Though  similar numerical results in the Schwarzschild case
were reported by Burko \cite{Burko},
the evolution of massive scalar fields at asymptotic late times 
has been claimed to be  inverse power-law behavior.
%, unlike \cite{HandP}.

Our purpose of this paper is to clarify what kind of 
mass-induced behaviors dominates in the asymptotic  late-time tails 
as a result of interaction of massive scalar fields with a black hole.  
In Sec. \ref{sec:Green's} we introduce 
the black-hole Green's function using the spectral decomposition method
\cite{Leaver}. (Another new method
to treat late-time tails 
which is called ``late time expansions''
was also proposed recently
\cite{Ori}.)
In Sec. \ref{sec:analysis} 
we consider a nearly extreme Reissner-Nordstr\"{o}m background,
motivated by the fact that the 
resonance phenomena in the vacuum polarization of quantum scalar fields
become clear.
Then, based on the procedure of asymptotic matching, 
we construct approximate solutions in the nearly extreme limit.
In Sec. \ref{sec:Intermediate} we study the intermediate tail which appears 
in the case of small mass field, 
confirming  our result in comparison with (\ref{eq:HP}).
In Sec. \ref{sec:Asymptotic} we derive the  asymptotic
late-time tail which is the main result
in this paper.
The final section is devoted to summary and discussion.

%%%%%%%%%%%%%%%%%  Green's function analysis  %%%%%%%%%%%%%%%%%%%%%%%%%%%%%%%%
\section{Green's function analysis}
\label{sec:Green's}
\subsection{Massive scalar fields in Reissner-Nordstr\"{o}m
geometry}
%Reissner-Nordstr\"{o}m$BGX7J;~6u>e$G$N8EE5E*(Bmassive $B%9%+%i!<>l$r9M$($k!#(B
%$B%a%H%j%C%/$O(B

We consider time evolution of a 
massive scalar field in Reissner-Nordstr\"{o}m
background with mass $M$ and charge $Q$.
The metric is
\begin{equation}
  \label{eq:metric}
  ds^2=-\left(1-\frac{2M}{r}+\frac{Q^2}{r^2}\right)dt^2
+\left(1-\frac{2M}{r}+\frac{Q^2}{r^2}\right)^{-1}dr^2
+r^2d\Omega ^2,
\end{equation}
%$B%9%+%i!<>l$,$_$?$94pK\J}Dx<0$O(B
and the scalar field $\phi$ with mass $m$ satisfies  the wave equation
\begin{equation}
\label{eq:wave}
\left(\Box -m^2\right)\phi =0.
\end{equation}
Resolving the field into spherical harmonics 
%$BJQ?tJ,N%$9$k(B
\begin{equation}
  \label{eq:separation}
  \phi = \sum _{\scriptscriptstyle l,m}
\frac{\psi _{\scriptscriptstyle l}(t,r)}{r}
Y_{\scriptscriptstyle l}^{\scriptscriptstyle m}(\theta ,\varphi),
\end{equation}
we obtain a wave equation for each multiple moment;
\begin{equation}
  \label{eq:radial}
  \psi _{,tt} -\psi _{,r_{\ast}r_{\ast}} +V \psi =0,
\end{equation}
where $r_{\ast}$ is the tortoise coordinate defined by
\begin{equation}
  \label{eq:tortoise}
  dr_{\ast}=\frac{dr}{1-\frac{2M}{r}+\frac{Q^2}{r^2}},
\end{equation}
and the potential $V$ is
\begin{equation}
  \label{eq:potential}
  V=\left(1-\frac{2M}{r}+\frac{Q^2}{r^2}\right)
\left[\frac{l(l+1)}{r^2}+\frac{2M}{r^3}-\frac{2Q^2}{r^4}\right].
\end{equation}

\subsection{The black-hole Green's function}
The time evolution of $\psi$ is given by
\begin{equation}
  \label{eq:evolution}
  \psi (r_{\ast},t)=\int \left[G(r_{\ast},r_{\ast}';t)\psi _t(r',0) 
+ G_t(r_{\ast},r_{\ast}';t)\psi (r_{\ast}',0) \right]
dr_{\ast}'
\end{equation}
for $t>0$, where $G(r_{\ast},r'_{\ast};t)$ is
the (retarded) Green's function satisfying 
\begin{equation}
  \label{eq:retarded}
  \left[\frac{\partial ^2}{\partial t^2} 
-\frac{\partial ^2}{\partial r_{\ast}^2} 
+V
\right]G(r_{\ast},r_{\ast}';t) = \delta (t)\delta (r_{\ast}-r_{\ast}')
\end{equation}
with the initial condition 
$G(r_{\ast},r'_{\ast};t)=0$ for $t\le 0$. 
We calculate 
the Green's function 
through  the Fourier transform;
%$BF07BItJ,(B$\psi _{\scriptscriptstyle \omega l}(t,r)$
%$B$N;~4VH/E8$O!"%0%j!<%s4X?t$+$i$o$+$k(B
\begin{equation}
  \label{inverse}
  G(r_{\ast},r_{\ast}';t)= -\frac{1}{2\pi}\int _{-\infty +ic}^{\infty +ic} 
\tilde{G}(r_{\ast},r'_{\ast};\omega)
e^{\scriptscriptstyle -i\omega t}d\omega ,
\end{equation}
where $c$ is some positive constant.
The usual procedure is to
close the contour of integration into the lower half of the 
complex frequency plane shown in Fig. \ref{fig1}. 
Then, the late-time tail behaviors which are our main concern should be
given by the integral along the branch cut
placed along the interval $-m \le \omega \le m$.

%%%%%%%%%%%%%%%%%%%%%%%%%%%%%%%%%%%%%%%%%%%%%%%%%%
%The Green's function 
The Fourier component $\tilde{G}(r_{\ast},r'_{\ast};\omega)$
in the range $-m \le \omega \le m$
 can be expressed in terms of two linearly independent solutions
$\tilde{\psi}_1$ and $\tilde{\psi}_2$
for the homogeneous equation
\begin{equation}
  \label{eq:homo}
  \left(\frac{d^2}{dr_{\ast}^2}+\omega ^2 -V\right)
\tilde{\psi}_i =0, \quad i=1,2.
\end{equation}
The boundary condition for the basic solution $\tilde{\psi}_1$
is to describe purely ingoing waves crossing the event horizon, i.e.,
\begin{equation}
  \label{eq:boundary1}
  \tilde{\psi}_1 \simeq e^{-i\omega r_{\ast}}, 
\end{equation}
as $r_{\ast} \to -\infty$,
while $\tilde{\psi}_2$ is required to damp exponentially
at spatial infinity, i.e.,
\begin{equation}
  \label{eq:boundary2}
  \tilde{\psi}_2 \simeq  e^{-\varpi r_{\ast}},
\end{equation}
as $r_{\ast} \to \infty$,
where $\varpi \equiv \sqrt{m^2-\omega ^2}$.
Because the complex conjugate $\tilde{\psi}_1^{\ast}$ is also a solution for 
Eq. (\ref{eq:homo}),
$\tilde{\psi}_2$ can be written by the linear superposition
\begin{equation}
  \label{eq:2to1}
  \tilde{\psi}_2 =
\alpha \tilde{\psi}_1 +\beta\tilde{\psi}_1^{\ast},
\end{equation}
and the Wronskian is estimated to be
%we obtain
\begin{equation}
  \label{}
W(\omega )=\tilde{\psi}_1\tilde{\psi}_{2,r_{\ast}}
-\tilde{\psi}_{1,r_{\ast}}\tilde{\psi}_2 =2i\omega \beta.
\end{equation}
%%%%%%%%%%%%%%%%%%%%%%%%%%%%%%%%%%%%%%%%%%%%%%%%%%%%%%%%%%%%%%%%
Using these two solutions, the Green's function can be written by
\begin{equation}
  \label{eq:Green}
 \tilde{G}(r_{\ast},r'_{\ast};\omega )= -\frac{1}{2i\omega \beta}
\left\{ \begin{array}{l@{\quad,\quad}l}
\tilde{\psi} _1(r'_{\ast},\omega)
\tilde{\psi }_2(r_{\ast},\omega)&
\qquad  r' >r ,\\
\tilde{\psi} _1(r_{\ast},\omega)
\tilde{\psi }_2(r'_{\ast},\omega)
&\qquad r'< r .
\end{array}
\right.
\end{equation}
The contribution $G^C$ from the branch cut to the Green's function
is reduced to 
\begin{eqnarray}
  \label{eq:branch-cut2}
  G^C (r_{\ast},r_{\ast}';t)&=&
-\frac 1{4\pi i}
\int _{\rm{cut}}
\frac 1\omega
\frac{\alpha}{\beta}
\tilde{\psi}_1(r_{\ast}',\omega)
\tilde{\psi}_1(r_{\ast},\omega)
e^{-i\omega t}d\omega .
%\nonumber\\
%&&+(\rm{complex} \> \rm{conjugate}).
\end{eqnarray}
%%%%%%%%%%%%%%%%%%%%%%%%%%%
%\begin{eqnarray}
%  \label{eq:branch-cut2}
%  G^C (r_{\ast},r_{\ast}';t)&=&
%-\frac 1{4\pi i}
%\int _{0}^m
%\frac 1\omega
%\tilde{\psi}_1(r_{\ast}',\omega)
%\tilde{\psi}_1(r_{\ast},\omega)
%\left[ \frac{\alpha(\omega ,\varpi)}{\beta(\omega ,\varpi)}
%-\frac{\alpha(\omega ,e^{-i\pi}\varpi)}{\beta(\omega ,e^{-i\pi}\varpi)}
%\right]e^{-i\omega t}d\omega \nonumber\\
%&&+(\rm{complex} \> \rm{conjugate}).
%\end{eqnarray}
Then the main task to evaluate $G^C$ is to derive the coefficients
$\alpha$ and $\beta$.

%%%%%%%%%%%%%%%%%%%%%% matching %%%%%%%%%%%%%%%%%%%%%%%%%%%%%%%%
\section{nearly extreme limit}
\label{sec:analysis}
It is difficult in general to obtain the  coefficients $\alpha$ and $\beta$,
since exact solutions for the wave equation (\ref{eq:homo}) cannot be 
expressed by any elementary function
or any trancedental function already known.
Fortunately, in nearly extreme Reissner-Nordstr\"{o}m geometry,
the procedure of asymptotic matching turn out to be very useful.
Let us change the variable $r$ 
%At first, we change the variable from $r$ 
to $z$ 
%which we define as
defined as
\begin{equation}
  \frac{z-1}{2}=\frac{r-r_+}{2 \kappa r_+^2},
\end{equation}
where $r_+$ and $r_-$ are the outer and inner horizon radii, respectively,
and $\kappa $ is the surface gravity defined as
$\kappa \equiv(r_+-r_-)/2r_+^2$.
Then we can rewrite   the wave equation (\ref{eq:homo}) into
\begin{eqnarray}
\label{eq:homoexact}
%\lefteqn{ 
(z^2-1)\frac{d^2\tilde{\psi}}{dz^2}
+\frac{ \left\{2\kappa r_+^2+z(r_++r_-)\right\}}
{r_+\{(z-1)\kappa r_++1\}}
\frac{d\tilde{\psi}}{dz}
%\nonumber\\&& 
+\Bigg[\frac{\omega ^2}{\kappa ^2}
\frac{\{(z-1)\kappa r_++1\}^4}{z^2-1}-l(l+1)&&
\nonumber\\
%&&
-m^2 r_+^2 \{(z-1)\kappa r_++1\}^2
-\frac{\kappa r_+^2\left\{2\kappa r_+^2 +z(r_++r_-)\right\}}
{ r_+^2 \{(z-1)\kappa r_++1\}^2}
\Bigg]\tilde{\psi}&=&0.
\end{eqnarray}
For the nearly extreme case such that
%$\kappa M\ll 1$,
\begin{equation}
  \label{eq:extremallimit}
  \kappa M\ll 1,
\end{equation}
we can allows us to derive the approximate solutions valid in the region
$z \gg 1$ or $z \ll 1/\kappa M$.
%%%%%%%%%%%%%%%%%%%%%%%%%%%%%%%%%%%%%%%%%%%%%%%%%%%%%%%%%%%%%%%%%%%%%%%%%%%%%%
\subsection{Solutions for $z\ll 1/\kappa M$}
%the region I ($z\ll 1/\kappa M$)}
%In the region $\kappa M z\ll 1$
%%Truncated $o(\kappa Mz)$ terms,
%We expand the wave equation (\ref{eq:homoexact}) as a power series 
%in $\kappa Mz$, truncate terms of order $O(\kappa M z)$ and higher,
%and obtain
Expanding the wave equation (\ref{eq:homoexact}) as a power series 
in $\kappa Mz$ and truncating  terms of order $ \kappa M z $ and higher,
we obtain
%is approximately given by
\begin{eqnarray}
\label{mode1}
  \lefteqn{(z^2-1)\frac{d^2\tilde{\psi}}{dz^2}
+2z 
%+o(\kappa Mz
\frac{d\tilde{\psi}}{dz}}\nonumber\\
&&+\left[\frac{1}{z^2-1}\frac{\omega ^2}{\kappa ^2}
+\frac{1}{z+1}
\left(\frac{4\omega ^2M}{\kappa}-12 \omega ^2M^2 
%+ o(\kappa Mz)
\right)
+\left\{6\omega ^2M^2-l(l+1)-m^2M^2
%+ o(\kappa Mz)
\right\}\right]\tilde{\psi}=0.
%\nonumber\\
\end{eqnarray}
%in the region $\kappa M z\ll 1$.
%We neglect $o(1)$ terms.
%Changing variables 
Then the solution $\tilde{\psi}_1$ satisfying the boundary condition 
(\ref{eq:boundary1})
can be written using the hypergeometric function $F$;
\begin{eqnarray}
  \label{eq:ingoing}
  \tilde{\psi}_1(\omega , r_{\ast})&=&
%\xi ^{(j_2 -j_3 -j_1 -1)/2}
\xi ^{\frac{i\omega}{\kappa}-2i\omega M +\frac 12 +\mu  }
%\left(\frac{\xi -1}{\xi}\right)^{(j_1 -j_2-j_3)/2}
\left(\xi -1\right)^{-\frac{i\omega}{2\kappa}}
\nonumber\\&&\times
F
\left(-\frac{i\omega}{\kappa} +2i\omega M -\mu +\frac 12 , 
-2i\omega M -\mu +\frac 12,-\frac{i\omega}{\kappa}+1 ; \frac{\xi -1}{\xi}
\right)\\
\label{eq:ingoing2}
&=& \xi ^{\frac{i\omega}{\kappa}-2i\omega M -\mu +\frac 12}
\left(\xi -1\right)^{-\frac{i\omega}{2\kappa} }
\frac{\Gamma(\frac{i\omega}{\kappa}+1 )\Gamma(-2\mu)}
{\Gamma(\frac{i\omega}{\kappa}-2i\omega M -\mu +\frac 12 )
\Gamma(2i\omega M -\mu +\frac 12)}\nonumber\\
&&\times 
F\left(2i\omega M +\mu +\frac 12,
\frac{i\omega}{\kappa}-2i\omega M +\mu +\frac 12,
2\mu +1;\frac 1\xi\right)\nonumber\\
&&+\xi ^{\frac{i\omega}{\kappa}-2i\omega M +\mu +\frac 12}
\left(\xi -1\right)^{-\frac{i\omega}{2\kappa} }
\frac{\Gamma(\frac{i\omega}{\kappa}+1 )\Gamma(2\mu)}
{\Gamma(\frac{i\omega}{\kappa}-2i\omega M +\mu +\frac 12)
\Gamma(2i\omega M +\mu +\frac 12)}\nonumber\\
&&\times 
F\left(2i\omega M -\mu +\frac 12,
\frac{i\omega}{\kappa}-2i\omega M -\mu +\frac 12,
-2\mu +1;\frac 1\xi\right),
\end{eqnarray}
%as $r_{\ast}\to -\infty$.
where new variable $\xi$ is defined as
\begin{equation}
  \xi \equiv \frac{z+1}{2},
\end{equation}
%and $j_1$, $j_2$ and $j_3$ are defined as
%\begin{equation}
%  j_1 =
%%1+2i\sqrt{\frac{\omega ^2}{4\kappa ^2}
%%-\frac{2\omega ^2M}{\kappa}+6\omega ^2M^2}.
%\frac{i\omega}{\kappa}-4i\omega M +1,
%\end{equation}
%\begin{equation}
%  j_2 =
%%i\sqrt{\frac{\omega ^2}{4\kappa ^2}
%%-\frac{2\omega ^2M}{\kappa}+6\omega ^2M^2}
%%+\frac{i \omega}{2\kappa}
%%+\frac{1}{2}
%%+\sqrt{\left(l+\frac{1}{2}\right)^2+m^2M^2-6 \omega ^2M^2},
%\frac{i\omega}{\kappa}-2i\omega M +\frac 12 +\mu ,
%\end{equation}
%\begin{equation}
%  j_3 =
%%i\sqrt{\frac{\omega ^2}{4\kappa ^2}
%%-\frac{2\omega ^2M}{\kappa}+6\omega ^2M^2}
%%+\frac{i \omega}{2\kappa}
%%+\frac{1}{2}
%%-\sqrt{\left(l+\frac{1}{2}\right)^2+m^2M^2-6 \omega ^2M^2},
%\frac{i\omega}{\kappa}-2i\omega M +\frac 12 -\mu 
%\end{equation}
and $\mu$ is 
\begin{equation}
  \mu \equiv \sqrt{\left(l+\frac{1}{2}\right)^2+m^2M^2-6 \omega ^2M^2}.
\end{equation}
We have used the linear transformation formulas 15.3.6 of \cite{Abramo}
in the second equality of (\ref{eq:ingoing}).
Using asymptotic expansions 15.7.2 and 15.7.3 of \cite{Abramo}
for $|i\omega /\kappa|\gg 1$,
we can reduce (\ref{eq:ingoing2}) to 
\begin{eqnarray}
  \label{zll1kM}
  \tilde{\psi}_1 (\omega , r_{\ast})
%&\simeq&\nonumber\\
&\simeq&
e^{\frac{i|\omega| }{\kappa z}}\left(\frac{2}{z}\right)^{2i|\omega| M}
e^{-\pi |\omega| M},
\end{eqnarray}
which is valid in the region $1 \ll z\ll 1/\kappa M$,
%and $\kappa M \ll 1$.
and will be used for asymptotic matching 
with the solutions given in $z \gg 1$.
%%%%%%%%%%%%%%%%%%%%%%%%%%%%%%%%%%%%%%%%%%%%%%%%%%%%%%%%%%%%%%%%%%%%%%%%%%%%%%%

%\newpage
\subsection{Solutions for $z \gg 1$}
%In the region $z \gg 1$, we expand 
%the wave equation (\ref{eq:homo}) as a power series in $1/z$,
%truncate terms of order $1/z$ and higher, and obtain
Expanding the wave equation (\ref{eq:homo}) as a power series in $1/z$
and truncating  terms of order $1/z$ and higher, we obtain
\begin{eqnarray}
\label{inner}
\lefteqn{\frac{d^2u}{dx^2}+\frac{2}{x}
\frac{du}{dx}}
\nonumber\\
&&
+\Bigg[
\omega ^2M^2-m^2M^2+\frac{1}{x}(4\omega ^2M^2-2m^2M^2)
+\frac{1}{x^2}\left\{6\omega ^2M^2-l(l+1)-m^2M^2\right\}\nonumber\\
&&+\frac{4\omega ^2M^2 }{x^3}+\frac{\omega ^2M^2 }{x^4}
+\frac{2}{x(x+1)^2}
\Bigg]u=0,
\end{eqnarray}
%is approximately given  by
%\begin{eqnarray}
%\lefteqn{\frac{d^2\tilde{\psi}}{dx^2}+\left(\frac{2}{x}-\frac{2}{x+1}\right)
%\frac{d\tilde{\psi}}{dx}}
%\nonumber\\
%&&
%+\Bigg[
%\omega ^2M^2-m^2M^2+\frac{1}{x}(4\omega ^2M^2-2m^2M^2)
%+\frac{1}{x^2}
%\left\{6\omega ^2M^2-l(l+1)-m^2M^2\right\}\nonumber\\
%&&+\frac{1}{x^3}4\omega ^2M^2+\frac{1}{x^4}\omega ^2M^2
%+\frac{2}{x(x+1)^2}
%\Bigg]\tilde{\psi}=0,
%\end{eqnarray}
introducing the new variable $x$ defined as 
%we employ as radial coordinate defined as 
\begin{eqnarray}
  x&\equiv&\kappa M z
\end{eqnarray}
and the function $u$ defined as 
\begin{eqnarray}
  \tilde{\psi} &\equiv& (x+1)u.
\end{eqnarray}
%$B$H$*$/$H(B $u$$B$KBP$9$kJ}Dx<0(B
We can give the approximate solutions for Eq. (\ref{inner}) 
using trancedental functions 
in the regions $x \ll 1$ and $x \gg 1$ respectively,
although it is difficult to find the exact solutions 
valid in the whole range of $x$.

%\newpage
%Now we divide the region $z \gg 1$ into two region $x \ll 1$ and $x \gg 1$
%$z \gg 1$$B$r$5$i$K#2$D$NNN0h$KJ,$1$k!#(B
%%%%%%%%%%%%%%%%%%%%%%%%%%%%%%%%%%%%%%%%%%%%%%%%%%%%%%%%%%%%%%%%%%%%%%%%%%%%%%%
\subsubsection{Solutions for $x \ll 1$}
Our strategy 
to find  the approximate solutions for Eq. (\ref{inner}) 
%using trancedental functions 
is to truncate terms of order $x^{-1}$ and higher in the coefficients of $u$,
since these can be smaller than other terms for $x \ll 1$,
%the  region I ($x \ll 1$)}
%Leaving the terms of order $x^{-4}$,  $x^{-3}$ and $x^{-2}$
%in the coefficients of $u$,  
%since          other terms can be neglected 
%in the region $x \ll 1$,
%Truncating terms of %which coefficients of $u$ is 
%order $x^{-1}$ and higher in the coefficients of $u$,
%In the region $x \ll 1$, 
and to
change the variable $x$ to 
\begin{eqnarray}
  s&=&
%2i\omega r_+ y=
\frac{2i\omega M}{x}.
%=\frac{2i\omega }{z\kappa}.
\end{eqnarray}
Then, %we can approximate 
Eq. (\ref{inner}) can be approximated by
%Eq.(\ref{inner})$B$O(B
%\begin{eqnarray}
%\label{inner1}
% \frac{d^2q}{dx^2}+\frac{2}{x}\frac{dq}{dx}
%+\Big[\frac{1}{x^4}\omega ^2M^2+\frac{1}{x^3}4\omega ^2M^2
%+\frac{1}{x^2}(6\omega ^2M^2-l(l+1)-m^2M^2)\Big]q&=&0 \nonumber\\ 
%\end{eqnarray}
%\begin{eqnarray*}
%  \frac{d^2\psi}{dx^2}+\left(\frac{2}{x}-\frac{2}{x+1}\right)\frac{d\psi}{dx}
%+\left[\frac{1}{x^4}\omega ^2r_+^2+\frac{1}{x^3}4\omega ^2r_+^2
%+\frac{1}{x^2}(6\omega ^2r_+^2-l(l+1)-m^2r_+^2)
%\right]\psi&=&0
%\end{eqnarray*}
%\begin{eqnarray*}
%  \psi &=&(x+1)u
%\end{eqnarray*}
%$B$H$*$/$H(B  
\begin{eqnarray}
\label{inner1}
   \frac{d^2u}{dx^2}+\frac{2}{x}\frac{du}{dx}
+\left[\frac{1}{x^4}\omega ^2M^2+\frac{1}{x^3}4\omega ^2M^2
+\frac{1}{x^2}
\left\{6\omega ^2M^2-l(l+1)-m^2M^2\right\}
\right]u&=&0, %\nonumber\\&&
\end{eqnarray}
%$B$G6a;w$G$-$k(B
and we can describe  
the solution of Eq. (\ref{inner1}) 
using the Whittaker's functions;
%is reduced to the Whittaker's functions;
%\begin{eqnarray}
%  \frac{dq^2}{ds^2}+\left[-\frac{1}{4}+\frac{\sigma_1 }{s}
%-\frac{\mu ^2-\frac14}{s^2}
%\right]=0,
%\end{eqnarray}
%and
%\begin{eqnarray}
%  \mu &=&\sqrt{\left(l+\frac{1}{2}\right)^2+m^2M^2-6\omega ^2M^2}
%\end{eqnarray}
%The solution can be written using Whittaker's functions;
\begin{eqnarray}
\label{xll1sol}
% \tilde{\psi}_2 
u
&=& a_1W_{\sigma_1 ,\mu}
\left(\frac{2i\omega M}{x}\right)
%(1+x)
+a_2W_{-\sigma_1,\mu}
\left(-\frac{2i\omega M}{x}\right),
%(1+x)
\end{eqnarray}
where
\begin{eqnarray}
\sigma_1&=&-2i\omega M.
\end{eqnarray}
The asymptotic expansions for 
$\displaystyle\left|\frac{2i\omega M}{x}\right| \gg 1$ lead to
%(\ref{xll1sol}) is reduced to
%We obtain 
\begin{eqnarray}
\label{xll1sol1asy}
 W_{\sigma_1,\mu}\left(\frac{2i\omega M}{x} \right)
  &\sim& 
e^{-i\frac{\omega}{\kappa z}}
\left(i\frac{2\omega }{\kappa z}\right)^{-2i\omega M},
\end{eqnarray}
which is necessary for asymptotic matching 
in the overlap region  with 
the solutions in the region $z \ll 1/\kappa M$.
On the other hand, using Eqs. 13.1.3, 13.1.4 and 13.1.33 of \cite{Abramo},
we can reduce Eq. (\ref{xll1sol})  to
\begin{eqnarray}
\label{xll1sol2}
 W_{\sigma_1,\mu}\left(\frac{2i\omega M}{x} \right)
%  &=& 
%\frac{\Gamma(-2\mu)}{\Gamma(\frac 12-\mu -\sigma_1)}
%M_{-2i\omega M,\mu}\left(\frac{2i\omega M}{x}\right)
%+\frac{\Gamma(2\mu)}{\Gamma(\frac 12+\mu -\sigma_1 )}
%M_{-2i\omega M,-\mu}\left(\frac{2i\omega M}{x}\right)\nonumber\\&& \nonumber\\
&\sim& 
\frac{\Gamma(-2\mu)}{\Gamma(\frac 12-\mu -\sigma_1)}
\left(\frac{2i\omega M}{x}\right)^{\mu +\frac 12}
+\frac{\Gamma(2\mu)}{\Gamma(\frac 12+\mu -\sigma_1)}
\left(\frac{2i\omega M}{x}\right)^{-\mu +\frac 12},
\end{eqnarray}
if the asymptotic expansions are applied in the region 
$\displaystyle\left|\frac{2i\omega M}{x}\right| \ll 1$
as an analytic extension.
%%%%%%%%%%%%%%%%%%%%%%%%%%%%%%%%%%%%%%%%%%%%%%%%%%%%%%%%%%%%%%%%%%%%%%%%%%%%%%%
\subsubsection{Solutions for $x \gg 1$}
%\subsubsection{region III  ($z \gg 1/\kappa M$)}
%In the region $x \gg 1$, 
Contrary to the region $x \ll 1$,
truncating terms of order $x^{-3}$ and higher in the coefficients of $u$
in Eq. (\ref{inner}),
since these can be smaller than other terms for $x \gg 1$,
%%leaving the terms of order $x^{0}$, $x^{-1}$ and $x^{-2}$
%truncating terms of 
%which coefficients of $u$ is 
%order $x^{-3}$ and higher 
%since other terms can be neglected in the region $x \gg 1$,
we can approximate Eq. (\ref{inner}) by
%\begin{eqnarray}
%\label{inner2}
%\lefteqn{  \frac{d^2q}{dx^2}+\frac{2}{x}\frac{dq}{dx}}\nonumber\\
%&&+\Big[\omega ^2r_+^2-m^2r_+^2+\frac{1}{x}(4\omega ^2r_+^2-2m^2r_+^2)
%+\frac{1}{x^2}(6\omega ^2r_+^2-l(l+1)-m^2r_+^2)\Big]q=0.  \nonumber\\
%\end{eqnarray}
\begin{eqnarray}
\label{inner2}
\lefteqn{\frac{d^2u}{dx^2}+\frac{2}{x}
\frac{du}{dx}}
\nonumber\\
&&
+\Bigg[
\omega ^2M^2-m^2M^2+\frac{1}{x}(4\omega ^2M^2-2m^2M^2)
+\frac{1}{x^2}
\left\{6\omega ^2M^2-l(l+1)-m^2M^2\right\}
\Bigg]u=0.
%\nonumber\\
\end{eqnarray}
Introducing the function $Z$ defined as 
\begin{eqnarray}
  u&=&\frac{Z}{x},
\end{eqnarray}
we can reduce Eq. (\ref{inner2}) into
%$Z$$B$KBP$9$kJ}Dx<0(B
\begin{eqnarray}
\label{z}
\frac{d^2Z}{dx^2}
+\Bigg[
\omega ^2M^2-m^2M^2+\frac{1}{x}(4\omega ^2M^2-2m^2M^2)
+\frac{1}{x^2}
\left\{6\omega ^2M^2-l(l+1)-m^2M^2\right\}
\Bigg]Z=0.
\end{eqnarray}
%\begin{eqnarray*}
%  q&=&\frac{W}{x}
%\end{eqnarray*}
%$B$9$k$H(B
%\begin{eqnarray*}
%\frac{d^2W}{dx^2}
%+\Big[\omega ^2M^2-m^2M^2+\frac{1}{x}(4\omega ^2M^2-2m^2M^2)
%+\frac{1}{x^2}(6\omega ^2M^2-l(l+1)-m^2M^2)\Big]W=0  \nonumber\\
%\end{eqnarray*}
%Eq. (\ref{z})$B$N2r$O(BWhittaker$B4X?t(B
Then we can write solutions for Eq. (\ref{z}) using Whittaker's functions;
%We choose $\tilde{\psi}_2$ as exponentially damping mode at spatial infinity;
\begin{eqnarray}
% \tilde{\psi}_2
u
&=&
%\underbrace{
b_1\frac{M_{\sigma_2,\mu}(2\varpi Mx )}{x}
+b_2\frac{M_{\sigma_2,-\mu}(2\varpi Mx )}{x},
%_{\rm{exponential\>\> damping\>\> mode}}
%\left(1+\frac 1x\right)
%\nonumber\\
%+B_2
%\underbrace{\left(1+\frac 1x\right)
%W_{-\tilde{\kappa_2},\mu}(-2\varpi r_+x )}
%_{\rm{exponential\>\> growing\>\> mode}}
%&\sim&e^{-\varpi Mx}(2\varpi Mx)^{\sigma_2}
\end{eqnarray}
%as $|2\varpi Mx| \gg 1$, 
where 
%\begin{eqnarray*}
%  \mu &=&\sqrt{\left(l+\frac{1}{2}\right)^2+m^2r_+^2-6\omega ^2r_+^2}
%\end{eqnarray*}
%(i)$m =0$$B$N>l9g(B
%\begin{eqnarray*}
%\tilde{\kappa_2}&=&-2i\omega r_+
%\end{eqnarray*}
%(ii)$m < \omega$$B$N>l9g(B
%\begin{eqnarray*}
%\tilde{\kappa}_2&=&-2i\sqrt{\omega ^2-m^2}r_+
%+\frac{m^2r_+}{i\sqrt{\omega ^2-m^2}}
%\end{eqnarray*}
%(iii)
%In the case of bound state ($m > \omega$)
\begin{eqnarray}
\sigma_2&=&-2\varpi M
+\frac{m^2M}{\varpi}.
\end{eqnarray}
%and %where
%\begin{eqnarray}
%  \varpi &\equiv& \sqrt{m^2-\omega ^2}.
%\end{eqnarray}
if estimated in the extended region $|2\varpi Mx| \ll 1$, 
we obtain
\begin{eqnarray}
\label{xgg1}
M_{\sigma_2,\mu}(2\varpi Mx )
%  W_{\sigma_2,\mu}(2\varpi Mx )
%%&=&
%%\frac{\Gamma(-2\mu)}{\Gamma(\frac 12 -\mu -\sigma_2 )}
%%M_{\tilde{\kappa} _2,\mu}(2\varpi Mx)
%%+\frac{\Gamma(2\mu)}{\Gamma(\frac 12 +\mu -\sigma_2 )}
%%M_{\tilde{\kappa} _2,-\mu}(2\varpi Mx)\nonumber\\&&\nonumber\\
&\sim&
%\frac{\Gamma (2\mu)(2\varpi Mx )^{-\mu+\frac{1}{2}}}
%{\Gamma(\frac{1}{2}+\mu -\sigma_2)}
%+\frac{\Gamma (-2\mu)(2\varpi Mx )^{\mu+\frac{1}{2}}}
%{\Gamma(\frac{1}{2}-\mu -\sigma_2)}
(2\varpi Mx)^{\mu +\frac 12},
\end{eqnarray}
using Eqs. 13.1.4 and 13.1.32 of \cite{Abramo}.
The solution $\tilde{\psi}_2$ satisfying the boundary condition 
(\ref{eq:boundary2}) is
\begin{eqnarray}
\label{boundphi2}
  \tilde{\psi}_2 &=& W_{\sigma_2,\mu}(2\varpi Mx )
%\nonumber\\
%&\sim&
\sim
e^{-\varpi Mx}(2\varpi Mx)^{\sigma_2}
\end{eqnarray}
for $|2\varpi Mx| \to \infty$.
%%%%%%%%%%%%%%%%%%%%%%%%%%%%%%%%%%%%%%%%%%%%%%%%
\subsection{Matching}
We can match 
both of asymptotic behaviors %of solutions
(\ref{zll1kM}) at  $z \ll 1/\kappa M$ and (\ref{xll1sol1asy}) at  $x \ll 1$
in the overlap region $1 \ll z\ll 1/(\kappa M)$, 
in order to determine the coefficients $a_1$ and $a_2$.
On the other hand,
%there are no  overlap regions between regions $x \ll 1$ and $x \gg 1$
%Nevertheless, 
we find that both of
asymptotic expressions (\ref{xll1sol2}) and (\ref{xgg1}),
that are the results due to analytic extensions from
one region into the other region each other,
have  similar forms.
So we can match these smoothly
in order to determine the coefficients $b_1$ and $b_2$.
In addition,
considering the boundary conditions
(\ref{eq:ingoing}) and (\ref{boundphi2})
imposed on 
$\tilde{\psi}_1$ and $\tilde{\psi}_2$, respectively,
 we can determine the coefficients $\alpha$ and $\beta$ 
in Eq. (\ref{eq:2to1}) as follows;
%As a  result we obtain
%For example, 
%if $\omega = |\omega|$ and $\varpi = |\varpi|$,
%$\alpha$ and $\beta$ reduce to
\begin{eqnarray}
\label{alpha}
  \alpha (|\omega| ,\varpi)&=&\beta (e^{i\pi}|\omega| ,\varpi)
\nonumber\\
&=&
\Bigg[\frac{\Gamma(2\mu)\Gamma(2\mu +1)(2\varpi M)^{-\mu+\frac 12}
e^{i \frac \pi 2(-\mu -\frac 12)}}
{\Gamma(\frac 12+\mu -\sigma _2)
(2|\omega| M)^{\mu+\frac{1}{2}}
\Gamma(\frac 12+\mu +2i|\omega| M)}\nonumber
\\&&%\nonumber\\&&
+\frac{\Gamma(-2\mu)\Gamma(-2\mu +1)(2\varpi M)^{\mu+\frac 12} 
e^{i \frac \pi 2(\mu -\frac 12)}}
{\Gamma(\frac 12-\mu -\sigma _2)
(2|\omega| M)^{-\mu+\frac{1}{2}}
\Gamma(\frac 12-\mu +2i|\omega| M)}
\Bigg]
\left(\frac{\omega}{\kappa}\right)^{2i|\omega| M}
e^{-\pi|\omega| M},
\end{eqnarray}
and
\begin{eqnarray}
\label{beta}
  \beta (|\omega| ,\varpi )&=&
\alpha(e^{i\pi}|\omega| ,\varpi)\nonumber\\
&=&
\Bigg[\frac{\Gamma(2\mu)\Gamma(2\mu +1)(2\varpi M)^{-\mu+\frac 12}
e^{i \frac \pi 2(\mu +\frac 12)}}
{\Gamma(\frac 12+\mu -\sigma _2)
(2|\omega| M)^{\mu+\frac{1}{2}}
\Gamma(\frac 12+\mu -2i|\omega| M)}\nonumber
\\&&%\nonumber\\&&
+\frac{\Gamma(-2\mu)\Gamma(-2\mu +1)(2\varpi M)^{\mu+\frac 12} 
e^{i \frac \pi 2(-\mu +\frac 12)}}
{\Gamma(\frac 12-\mu -\sigma _2)
(2|\omega| M)^{-\mu+\frac{1}{2}}
\Gamma(\frac 12-\mu -2i|\omega| M)}
\Bigg]
\left(\frac{|\omega|}{\kappa}\right)^{-2i|\omega| M}
e^{-\pi|\omega| M}.
\end{eqnarray}
\section{Intermediate late-time tails}  
\label{sec:Intermediate}
We consider the late-time behaviors of $G^C$ at the timescale
\begin{equation}
  \label{eq:late-time}
  mt \gg 1,
\end{equation}
when the decaying tails will dominate.
Hod and Piran \cite{HandP} pointed out that 
for the scalar field with small mass, namely $mM\ll 1$,
%\begin{equation}
%  \label{eq:small_mass}
%  mM \ll 1,
%\end{equation}   
%intermediate late-time tails (\ref{eq:HP}) dominate 
the dominant behavior is given by Eq. (\ref{eq:HP})
at the intermediate late times in the range
\begin{equation}
  \label{eq:time_window}
  mM \ll mt \ll \frac{1}{(mM)^2}.
\end{equation}
%by numerical simulations.
In this section
we check the validity of Eqs. (\ref{alpha}) and (\ref{beta}),
by deriving the intermediate tail behaviors.
Following Hod and Piran \cite{HandP},
the effective contribution to the integral in Eq. (\ref{eq:branch-cut2})
is claimed to be limited to the range $|\omega - m| =O(1/t)$
%arises from $|\omega | =O(m-1/t)$ 
or equivalently 
$\varpi = O(\sqrt{m/t})$.
This is due to the rapidly oscillating term $e^{-i\omega t}$
which leads to a mutual cancellation between the positive and the negative 
parts of the integrand.
Then, in the time scale given by Eq. (\ref{eq:time_window}) 
we note that the frequency range
$\varpi =O(\sqrt{m/t})$ leads to the inequality
\begin{equation}
\label{rela:intermediate}
%  \frac{m^2M}{\varpi}\ll 1
\sigma _2 \ll 1.
\end{equation}  
The factor $\sigma _2$, 
including field's parameter $m$ coupled with spacetime parameter $M$,
originates from the terms of order $x^{-1}$ 
in the coefficients of $u$ in 
Eqs. (\ref{inner}) or (\ref{inner2}).
If the relation (\ref{rela:intermediate}) is satisfied,
%the dominant contribution to the integral within 
%Eq. (\ref{rela:intermediate}),
%backscattering from the spacetime curvature has not dominate yet.
%the term is still smaller than other terms,
wave equation at far region can be approximated by
that of flat spacetime, which means that 
the effects
of backscattering due to spacetime curvature has not dominate yet.
In other words,
%That is, 
the value of $\sigma _2$ which gives effective contributions
to the integral (\ref{eq:branch-cut2})
represents a degree of the domination of the backscattering.

The relation (\ref{rela:intermediate})
 allows us to obtain the approximations of $\alpha$ and $\beta$
as follows,
\begin{eqnarray}
    \alpha (|\omega|,\varpi )
&\sim&
\Bigg[\frac{\Gamma(2l+1+2\epsilon _{\mu})
\Gamma(2l+2+2\epsilon _{\mu})}
{\Gamma(l+1+\epsilon _{\mu}-\sigma _2)
\Gamma(l+1+\epsilon _{\mu}+2i|\omega| M)}
(2\varpi M)^{-l-\epsilon _{\mu}}
(2|\omega| M)^{-l-1-\epsilon _{\mu}}
e^{i \frac \pi 2 (-l-1-\epsilon _{\mu}) }
\nonumber\\&&\nonumber\\&&
+\frac{\Gamma(-2l-1-2\epsilon _{\mu})
\Gamma(-2l-2\epsilon _{\mu})}
{\Gamma(-l-\epsilon _{\mu}-\sigma _2)
\Gamma(-l-\epsilon _{\mu}+2i|\omega| M)}
(2\varpi M)^{l+1+\epsilon _{\mu}} 
(2|\omega| M)^{l+\epsilon _{\mu}}
e^{i \frac \pi 2 (l+\epsilon _{\mu})}
\Bigg]\nonumber\\&&\nonumber\\&&
\times \left(\frac{|\omega|}{\kappa}\right)^{2i|\omega| M}
e^{-|\omega| M},
\end{eqnarray}
and
\begin{eqnarray}
  \beta(|\omega|,\varpi)
&\sim&
\Bigg[\frac{\Gamma(2l+1+2\epsilon _{\mu})
\Gamma(2l+2+2\epsilon _{\mu})}
{\Gamma(l+1+\epsilon _{\mu}-\sigma _2)
\Gamma(l+1+\epsilon _{\mu}-2i|\omega| M )}
(2\varpi M)^{-l-\epsilon _{\mu}}
(2|\omega| M)^{-l-1-\epsilon _{\mu}}
e^{i \frac \pi 2 (l+1+\epsilon _{\mu}) }
\nonumber\\&&\nonumber\\&&
+\frac{\Gamma(-2l-1-2\epsilon _{\mu})
\Gamma(-2l-2\epsilon _{\mu})}
{\Gamma(-l-\epsilon _{\mu}-\sigma _2)
\Gamma(-l-\epsilon _{\mu}-2i|\omega| M)}
(2\varpi M)^{l+1+\epsilon _{\mu}} 
(2|\omega| M)^{l+\epsilon _{\mu}}
e^{i \frac \pi 2 (-l-\epsilon _{\mu})}
\Bigg]\nonumber\\&&\nonumber\\&&
\times\left(\frac{|\omega|}{\kappa}\right)^{-2i|\omega| M}
e^{-|\omega| M},
\end{eqnarray}
where 
\begin{equation}
  \epsilon _{\mu}\equiv
\mu -\left(l+\frac 12\right) \simeq O(mM) \ll 1.
%\simeq  -\frac{5m^2M^2}{(2l+1)}.
\end{equation}
%$\epsilon _{\mu}=-\frac{5m^2M^2}{(2l+1)}$
%%%%%%%%%%%%%%%%%%%%%%%%%%%%%%%%%%%%%%%%%%%%%%%%%%%%%%%%%%%%%%%%%%%%%%%%%%%%%%
%Using the formula
%\begin{equation}
  %\end{equation}
Expanding the ratio $\alpha /\beta $ as a power series in $mM$, 
we can apploximate it as follows;
\begin{eqnarray}
\label{eq:inter-subtract}
  \frac{\alpha (|\omega| ,\varpi)}{\beta(|\omega|,\varpi)}
-\frac{\alpha (|\omega| ,e^{i\pi}\varpi)}
{\beta(|\omega| ,e^{i\pi}\varpi )}
&\sim &
\frac{l!^4}{(2l)!^2(2l+1)!^2}
(2M)^{4l+2} |\omega |^{2l+1}
\varpi ^{2l+1}
2i.
\end{eqnarray}
%in this time zone.  
%Approximating the integrand function
%using the relation (\ref{rela:intermediate}),
%The branch cut contribution from the Green's function 
Substituting Eq. (\ref{eq:inter-subtract}) into 
Eq. (\ref{eq:branch-cut2}), we obtain
%is reduced to
\begin{eqnarray}
\label{eq:interHandP}
G^C (r_{\ast},r'_{\ast};t) 
&= &
\frac{l!^4}{(2l)!^2(2l+1)!^2}
(2M)^{4l+2}
\int 
_{0}^{m}
\omega  ^{2l+1}
\varpi ^{2l+1}
%2
%\left[
\tilde{\psi}_1(r_{\ast},\omega)\tilde{\psi}_1(r'_{\ast},\omega)
e^{-i\omega t}
%+\tilde{\psi}_1(r_{\ast},\omega)\tilde{\psi}_1(r'_{\ast},\omega)
%e^{i\omega t}
%\right]
%\tilde{\psi}_1(r_{\ast},\omega)\tilde{\psi}_1(r'_{\ast},\omega)
%\cos(\omega t)
%e^{-i\omega t}
d\omega\nonumber\\&&
+ ({\rm complex\> conjugate}),
\end{eqnarray} 
which is similar to Eq. (29) in \cite{HandP},
giving the 
damping exponent in Eq. (\ref{eq:HP}).
%at the intermediate late time. 

Different from \cite{HandP},
our analytical calculation is not based on the 
flat space approximation.
The intermediate tails dominate in the range (\ref{eq:time_window}), 
when the integrand can be  approximated by Eq. (\ref{rela:intermediate}). 
It is easy to  find that the larger the field's mass is, 
the sooner it leaves the intermediate tails, and 
the phase does not appear in the case of $mM \gtrsim 1$.
%Futhermore we can find the behavior after the intermediate late times
%from our formalism.  

%%but to construct each mode by requiring boundary conditions.
%and it has advantage of estimating the following things
%analytically;
%The time scale when intermediate late-time tails dominate 
%can be estimated as (\ref{eq:time_window}) which is equivalent
%with that $\tilde{G}$ can be approximated using (\ref{rela:intermediate}).
%%by our formalism.
%That is, the larger the field's mass is, 
%the sooner it leaves the intermediate tails, and 
%the phase does not appear in the case of $mM \gtrsim 1$.
%%These agree to their numerical simulations \cite{HandP}.

%Intermediate tails depend only on the field's parameters 
%(namely, on the mass of the field) and it do not depend on the spacetime
%parameters.

%%%%%%%%%%%%%%%%%%%%% Asymptotic late-time tails %%%%%%%%%%%%%%%%%%%%%%%%%%%%%%
%\newpage
\section{Asymptotic late-time tails}
\label{sec:Asymptotic}
\begin{figure}
%\begin{center}
\begin{center}
\leavevmode
\epsfxsize=80mm
\epsfbox{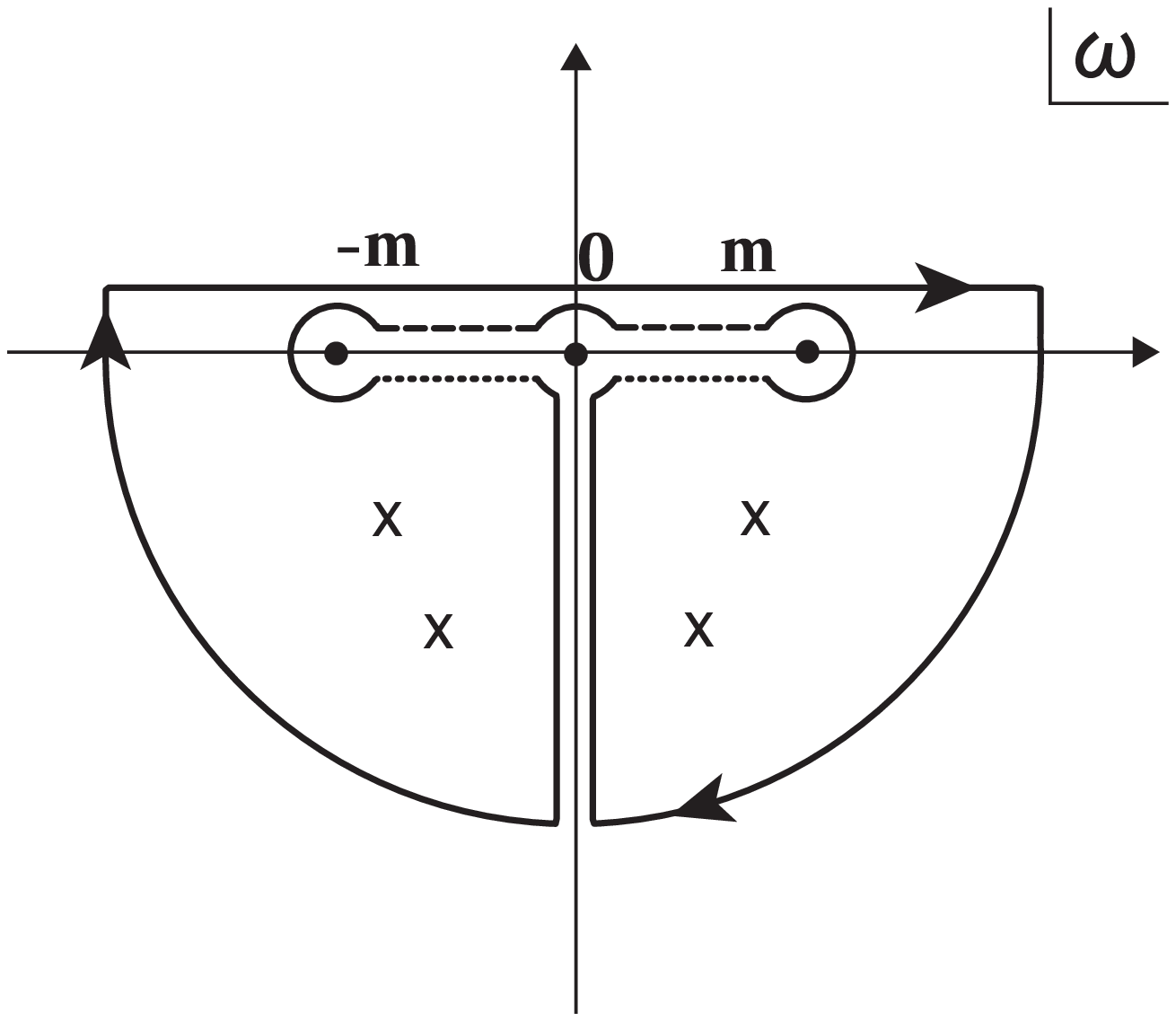}  
%\end{center}       
\caption{Integration contours in the complex frequency plane.
The original integration contour for the Green's function lies above 
the real ferquency axis.
We choose the value of $\varpi$ on the dashed line to be  $\varpi = |\varpi|$
and that on the dotted lines to be  $\varpi =e^{\pm i\pi}|\varpi|$.
The poles in $\tilde{G}(r_{\ast},r_{\ast }';\omega)$ are also shown, 
which give the quasinormal modes.
%When analytically continued in the complex plane
%this contour can be replaced by 
%the sum of high frequency arcs, the quasinormal modes,
%and an integral along the branch cut
%that leads to late-time tails
}
\label{fig1}
\end{center}
\end{figure}

It is obvious from our calculation
that the intermediate tail is  not a final pattern of decay
but should be replaced by another one,
because 
the dominant contribution to the integral 
(\ref{eq:branch-cut2}) is 
%made from the region 
out of the region (\ref{rela:intermediate}) 
after the intermediate late times (\ref{eq:time_window}).
%after the intermediate late times 
%(\ref{eq:time_window}).
The change into another phase 
was also numerically suggested in \cite{HandP}.
Physically the change of the tail behavior will be a result of dominant 
backscattering due to spacetime curvature,
which is the effect beyond the flat space approximation.
What kind of wave pattern dominates at very late times? 
In addition, 
we must reveal 
late-time tails in the $mM \gtrsim 1$ case of large field mass,
for which the intermediate tails do not appear.
In this section we study a tail behavior dominant
at asymptotic late times
%We consider very 
%late times after intermediate late time (\ref{eq:time_window})
\begin{equation}
  \label{saddle1}
mt \gg \frac{1}{m^2M^2},
\end{equation}
%%when for the parameter $\sigma _2$ we obtain
when the effective contribution to the integral 
(\ref{eq:branch-cut2}) arises from
 the region 
%  At that time the following relation
\begin{equation}
\label{rela:asymptotic}
%  \frac{m^2M}{\varpi}\gg 1
\sigma _2 \simeq \frac{m^2M}{\varpi}\gg 1,
\end{equation}
different from the inequality (\ref{rela:intermediate}) 
at intermediate late times.
%We mention that 
%due to spacetime curvature.
%In other words,
%the backscattering from spacetime curvature dominate,
%as the effective contribution to the integral
%arises 
%from the range (\ref{rela:asymptotic}).
%Approximating $\tilde{G}$ using the relation 
%(\ref{rela:asymptotic}),  
%Then, 
The coefficients 
$\alpha (\omega ,e^{\pm i\pi}\varpi)$ and $\beta (\omega ,e^{\pm i\pi}\varpi)$
are approximated 
for the inequality (\ref{rela:asymptotic}) 
by
\begin{eqnarray}
  \alpha (|\omega|,e^{-i\pi}\varpi)
&=&
\beta (e^{i\pi}|\omega|,e^{i\pi}\varpi)\nonumber\\
&\simeq &
\frac{1}{\sqrt{2\pi}}
e^{\frac{m^2M}{\varpi}}
(2m^2M^2)^{-\frac{m^2M}{\varpi}+2\varpi M}
(2\varpi M)^{ -2\varpi M+\frac{m^2M}{\varpi}+\frac 12}\nonumber\\&&
\Bigg[\frac{\Gamma(2\mu)\Gamma(2\mu +1)(2m^2M^2)^{\mu}
(2|\omega| M)^{-\mu-\frac{1}{2}}
e^{i \frac \pi 2 (\mu -\frac 32) }}
{\Gamma(\frac 12+\mu +2i|\omega| M)}
\nonumber\\&&
+\frac{\Gamma(-2\mu)\Gamma(-2\mu +1)(2m^2M^2)^{-\mu}
(2|\omega| M)^{\mu-\frac{1}{2}}
e^{i \frac \pi 2 (-\mu -\frac 32)}}
{\Gamma(\frac 12-\mu +2i|\omega| M)}
\Bigg]
\left(\frac{|\omega|}{\kappa}\right)^{2i|\omega| M}
e^{-\pi|\omega| M},
\end{eqnarray}
and
\begin{eqnarray}
    \beta(|\omega|,e^{-i\pi}\varpi)
&=& \alpha (e^{i\pi} |\omega|,e^{i\pi}\varpi) \nonumber\\
&\simeq &
\frac{1}{\sqrt{2\pi}}
e^{\frac{m^2M}{\varpi}}
(2m^2M^2)^{-\frac{m^2M}{\varpi}+2\varpi M}
(2\varpi M)^{ -2\varpi M+\frac{m^2M}{\varpi}+\frac 12}\nonumber\\&&
\Bigg[
\frac{\Gamma(2\mu)\Gamma(2\mu+1)
(2m^2M^2)^{\mu}
(2|\omega| M)^{-\mu -\frac 12}
e^{i \frac \pi 2(3\mu -\frac 12)}}
{\Gamma(\frac 12+\mu -2i|\omega| M)}
\nonumber\\&&
+
\frac{\Gamma(-2\mu)\Gamma(-2\mu+1)
(2m^2M^2)^{-\mu}
(2|\omega| M)^{\mu -\frac 12}
e^{i \frac \pi 2(-3\mu -\frac 12)}}
{\Gamma(\frac 12-\mu -2i|\omega| M)}
\Bigg]
\left(\frac{|\omega|}{\kappa}\right)^{-2i|\omega| M}
e^{-\pi|\omega| M}.
\end{eqnarray}
%because of Eq. (\ref{rela:asymptotic}). 
Thus the contribution from this part 
corresponding to the dotted lines in Fig. \ref{fig1} to 
the Green's function  given by
\begin{eqnarray}
\label{t-1tail}
  \frac{1}{4\pi i}
\int _{\rm{dotted\> lines}}
\frac{1}{\omega}
%\left[
\frac{\alpha (\omega ,\varpi)}
{\beta(\omega ,\varpi )}
\tilde{\psi}_1(r_{\ast},\omega)\tilde{\psi}_1(r'_{\ast},\omega)
e^{-i\omega t}d\omega
\end{eqnarray}
is $O(t^{-1})$ at most, 
because 
$\alpha(\omega ,e^{\pm i\pi}\varpi)$ and $\beta(\omega ,e^{\pm i\pi}\varpi)$
converge in the limit $|\omega|\to m$.
On the other hand,
%we find that the main contributions to the integral path 
%%integrand function $\tilde{G}(r,r';\omega)$ 
%are divided into two parts; 
%$\varpi $
%and $e^{\pm i\pi } \varpi$
%according to with or without rapidly oscillating terms
%in $\tilde{G}(r,r';\omega)$.
%The contributions from path along $\varpi $
%is important for 
%asymptotic late-time tail. 
%In this case (\ref{rela:asymptotic}),
$\alpha (\omega,\varpi)$ and $\beta (\omega,\varpi)$
%in (\ref{eq:branch-cut2})
are  reduced to
%We obtain
%We write down
\begin{equation}
  \frac{\alpha (|\omega|,\varpi)}{\beta(|\omega|,\varpi)}
=\frac{\beta(e^{i\pi}|\omega|,\varpi)}{\alpha (e^{i\pi}|\omega|,\varpi)}
\simeq \frac
{\gamma ^{\ast}e^{-i\pi \sigma _2}
+\eta ^{\ast}e^{i\pi \sigma _2}}
{\gamma e^{i\pi \sigma _2}
+\eta e^{-i\pi \sigma _2}},
\end{equation}
%%%%%%%%%%%%%%%%%%%%%%%%%%%
%%%%%%%%%%%%%%%%%%%%%%%%%%%
%and
%\begin{equation}
%  \frac{\alpha (e^{\pi i}\omega,e^{2\pi i}\varpi)}
%{\alpha ^{\ast}(e^{\pi i}\omega,e^{2\pi i}\varpi)}
%\simeq \frac
%{\gamma e^{i\pi \frac{m^2M}{\varpi}}
%+\eta e^{-i\pi \frac{m^2M}{\varpi}}}
%{\gamma ^{\ast}e^{-i\pi \frac{m^2M}{\varpi}}
%+\eta ^{\ast}e^{i\pi \frac{m^2M}{\varpi}}},
%\end{equation}
%since $\frac{m^2M}{\varpi} \gg 1$
for the limit  (\ref{rela:asymptotic}), where
\begin{eqnarray}
  \gamma &=&\frac{\Gamma(-2\mu)\Gamma(-2\mu+1)
(2m^2M^2)^{-\mu}
(2|\omega| M)^{\mu -\frac 12}}
{\Gamma(\frac 12-\mu -2i|\omega|M)}
e^{i\frac {\pi}{2}(\mu +\frac 12)}
\nonumber\\&&
+
\frac{\Gamma(2\mu)\Gamma(2\mu+1)(2m^2M^2)^{\mu}
(2|\omega| M)^{-\mu -\frac 12}}
{\Gamma(\frac 12+\mu -2i|\omega|M )}
e^{i\frac {\pi}{2}(-\mu +\frac 12)},
\end{eqnarray}
and
\begin{eqnarray}
  \eta &=&\frac{\Gamma(-2\mu)\Gamma(-2\mu+1)
(2m^2M^2)^{-\mu}
(2|\omega| M)^{\mu -\frac 12}}
{\Gamma(\frac 12-\mu -2i|\omega|M )}
e^{i\frac {\pi}{2}(-3\mu +\frac 12)}
\nonumber\\&&
+
\frac{\Gamma(2\mu)\Gamma(2\mu+1)(2m^2M^2)^{\mu}
(2|\omega| M)^{-\mu -\frac 12}}
{\Gamma(\frac 12+\mu -2i|\omega|M)}
e^{i\frac {\pi}{2}(3\mu +\frac 12)}.
\end{eqnarray}
%and
%\begin{eqnarray}
%c&=&
%\frac{\Gamma(-2\mu)\Gamma(-2\mu+1)
%(2m^2M^2)^{-\mu}
%(2|\omega| M)^{\mu -\frac 12}}
%{\Gamma(\frac 12-\mu -2i|\omega|M)}
%e^{i\frac {\pi}{2}(-\mu +\frac 12)}
%\nonumber\\&&\nonumber\\
%d&=&
%\frac{\Gamma(2\mu)\Gamma(2\mu+1)(2m^2M^2)^{\mu}
%(2|\omega| M)^{-\mu -\frac 12}}
%{\Gamma(\frac 12+\mu -2i|\omega|M)}
%e^{i\frac {\pi}{2}(\mu +\frac 12)}
%\end{eqnarray}
%The branch cut contribution to the Green's function is
Therefore we find the contribution from this part 
corresponding to the dashed line in Fig. \ref{fig1} to 
the Green's function to be approximated by
\begin{eqnarray}
\label{asympto-int}
%G^{C}(r_{\ast},r'_{\ast};\omega) 
%&\simeq &
&&
\frac{1}{4\pi i}
\int _0^m
\frac{1}{\omega}
%\left[
\frac{\alpha (\omega ,\varpi )}
{\beta(\omega ,\varpi )}e^{-i\omega t}
\tilde{\psi}_1(r_{\ast},\omega)\tilde{\psi}_1(r'_{\ast},\omega)
%\right.
%\nonumber\\&&
%\left.
%-\frac{\alpha (e^{2\pi i}\varpi ,e^{\pi i}\omega)}
%\alpha ^{\ast}(e^{2\pi i}\varpi ,e^{\pi i}\omega)}e^{i\omega t}
%\tilde{\psi}_1(r_{\ast},e^{\pi i}\omega)
%\tilde{\psi}_1(r'_{\ast},e^{\pi i}\omega)
%\right]
d\omega 
%&=&\int 
%\frac{1}{2i\omega}
%e^{i(\phi -\omega t)}
+({\rm complex \> conjugate })
\nonumber\\
&\simeq&
\frac{1}{4\pi mi}
\tilde{\psi}_1(r_{\ast},m)
\tilde{\psi}_1(r'_{\ast},m)
\int _0^m
%\frac{1}{\omega}
%\left[
e^{i(2\pi\sigma _2 -\omega t)} 
e^{i\varphi}
%-e^{i(-2\pi\sigma _2 +\omega t)} 
%e^{-i\varphi}
%\tilde{\psi}_1^{\ast}(r_{\ast},\omega)
%\tilde{\psi}_1^{\ast}(r'_{\ast},\omega)
%\right]
d\omega +({\rm complex \> conjugate }),
\end{eqnarray}
%%%%%%%%%%%%%%%%%%%%%%%%%%%%%%
%\begin{eqnarray}
%G^{C}(r_{\ast},r'_{\ast};\omega) &\sim &\int 
%\frac{1}{2i\omega}
%\left[
%\frac{\alpha (\varpi ,\omega)}
%{\alpha ^{\ast}(\varpi ,\omega)}e^{-i\omega t}
%\tilde{\psi}_1(r_{\ast},\omega)\tilde{\psi}_1(r'_{\ast},\omega)
%\right.\nonumber\\&&
%\left.
%-\frac{\alpha (e^{2\pi i}\varpi ,e^{\pi i}\omega)}
%{\alpha ^{\ast}(e^{2\pi i}\varpi ,e^{\pi i}\omega)}e^{i\omega t}
%\tilde{\psi}_1(r_{\ast},e^{\pi i}\omega)
%\tilde{\psi}_1(r'_{\ast},e^{\pi i}\omega)
%\right]d\omega 
%&=&\int 
%\frac{1}{2i\omega}
%e^{i(\phi -\omega t)}
%\nonumber\\
%&=&
%\int 
%\frac{1}{2i\omega}
%\left[
%e^{i(2\pi\tilde{\kappa}_2 -\omega t)} 
%e^{i\varphi}
%\tilde{\psi}_1(r_{\ast},\omega)
%\tilde{\psi}_1(r'_{\ast},\omega)
%-e^{i(-2\pi\frac{m^2M}{\varpi} +\omega t)} 
%e^{-i\varphi}
%\tilde{\psi}_1^{\ast}(r_{\ast},\omega)
%\tilde{\psi}_1^{\ast}(r'_{\ast},\omega)
%\right]
%d\omega
%\end{eqnarray}
%%%%%%%%%%%%%%%%%%%%%%%%%%%%%%%%%%%
where the phase $\varphi$ is defined by
\begin{equation}
  e^{i\varphi}=\frac{\eta ^{\ast}
+\gamma ^{\ast}e^{-2i\pi \sigma _2}}
{\eta +\gamma e^{2i\pi \sigma _2}}
\end{equation}
and it remains in the range $0 \le\varphi \le 2\pi$, 
even if $\sigma _2$ becomes very large,
since we have
\begin{equation}
|\eta |^2-|\gamma|^2
%=\eta \eta ^{\ast}- \gamma ^{\ast}\gamma 
=\frac{\pi}{|\omega| M}e^{2\pi |\omega| M}>0.
\end{equation}
%If conditions
%(\ref{eq:late-time}) and 
%\begin{equation}
%  \label{saddle1}
%mt \gg \frac{1}{m^2M^2}
%\end{equation}
%and
At very late times when 
(\ref{eq:late-time}) and (\ref{rela:asymptotic}) are satisfied,
%the the condition 
%the integration (\ref{eq:branch-cut2}) is dominated by
%the region both (\ref{eq:late-time}) 
%and (\ref{rela:asymptotic}) are satisfied,
both terms of $e^{i\omega t}$ and $e^{2i\pi \sigma _2}$ are 
rapidly oscillating.
In physical meaning, 
%At that time, 
scalar waves are mixed states with multiple phases 
backscattered by 
spacetime curvature,
and most of these waves are canceled out each other
by those of the inverse phase.
If
%there exists the frequency $\omega $ near $m$
the value of $2\pi \sigma _2 -\omega t$ in (\ref{asympto-int})
is stationary at $\omega =\omega _0$, i.e.,
\begin{equation}
\label{eq:saddle}
  \frac{d}{d\omega}\left( 2\pi \sigma _2 -\omega t\right)
%\Bigg 
%| _{\omega =\omega _0}
=0,
\end{equation}
%%however,
%and the value of $\omega _0$ is in the narrow region near $m$ moreover,
%then $\omega _0$ is a saddle point,
%of the function $2\pi \sigma _2 -\omega t$.
%As a result, 
%which means 
particular waves 
with  the frequency  $\omega _0$
% without cancellation, 
%satisfies Eq. (\ref{eq:saddle})
%\begin{equation}
%\label{eq:saddle}
%  \frac{d}{d\omega}\left( 2\pi \sigma _2 -\omega t\right)
%%\Bigg 
%%| _{\omega =\omega _0}
%=0,
%\end{equation}
remain without cancellation, and contribute dominantly
to the 
%very late-time 
tail behaviors. 
In such a case we can evaluate 
the integral (\ref{asympto-int})
as the effective contribution from
the immediate vicinity of the saddle point $\omega _0$.
This method which is called the saddle-point integration
allows us to evaluate accurately
the asymptotic behaviors of Bessel functions               
as a well-known example.
We can find a solution for Eq. (\ref{eq:saddle})
\begin{equation}
\label{saddlepoint}
  \varpi _0 \equiv \sqrt{m^2-\omega _0 ^2 }
\simeq \left(\frac{2\pi m^3 M}{t}\right)^{\frac 13}
\end{equation}
in the limit $\omega _0 \simeq m$.
In order for the  saddle point (\ref{saddlepoint}) 
to exist in the region (\ref{rela:asymptotic}),
where $e^{2i\pi \sigma _2}$ are 
rapidly oscillating as a function of $\omega$, 
we need the additional relation
\begin{equation}
  \label{saddle2}
mt \gg  mM.
\end{equation}
Approximating  the integration (\ref{asympto-int}) 
by the contribution from the immediate vicinity of $\omega _0$,
we obtain
\begin{eqnarray}
  \label{saddle-int}
  \frac{1}{4\pi im}
\int e^{i \frac{d^2}{d\omega^2}
\left( 2\pi \sigma _2 -\omega t \right)
\big|_{\omega =\omega _0}(\omega -\omega _0)^2}
e^{i\varphi (\omega_0)}d\omega 
&\sim&
\frac i{4\sqrt{3}}(2\pi)^{\frac 56}
(mM)^{\frac 13}(mt)^{-\frac 56}
e^{imt}
%e^{i 2(2\pi mM)^{\frac 23}(mt)^{\frac 13}}
e^{i\varphi (\omega_0)},
\end{eqnarray}
through the formula
%In the derivation of (\ref{saddle-int}),  we use the formula;
\begin{equation}
  \label{eq:frenel}
  \int _{-\infty}^{\infty}\cos (x^2) dx 
= \int _{-\infty}^{\infty}\sin (x^2) dx 
=\sqrt{\frac{\pi}{2}}.
\end{equation}
%and $\varphi _0 =\varphi(\omega _0)$.
%%%The integrand function $\tilde{G}e^{-i\omega t}$ contains
%%%rapidly oscillating terms
%%%$e^{-i\omega t}$ and $e^{\pm \frac{i\pi m^2M}{\varpi}}$.
%%%In addition to late-time condition (\ref{eq:late-time})
%%%%$mt \gg 1$,
%%%if the following conditions
%%%\begin{equation}
%%%  \label{saddle1}
%%%mt \gg \frac{1}{m^2M^2}
%%%\end{equation}
%%%\begin{equation}
%%%  \label{saddle2}
%%%mt \gg  mM  
%%%\end{equation}
%%%are satisfied,
%there exist saddle points.
%we can evaluate this integrand by the method of steepest descent.
%The saddle points are located at 
%$\omega = \omega _0 = 
%\pm m \sqrt{1-\left(\frac{2\pi M}{t}\right)^{\frac 23}}
%\sim m \left\{1-\frac{1}{2}\left(\frac{2\pi M}{t}\right)^{\frac 23} \right\}$
%$B;D$k$N$O(B $B<!$N#3$D$N>r7o$,B7$C$?>l9g$G$"$k!#(B
%$B0lC6$3$l$i$N>r7o$,$=$m$&$H(B $B$=$N8e$b$:$C$HB7$&(B
%$B$N$G!"(Basymptotic $B$G$"$k$H$$$($k(B 
%branch cut $B$+$i$N4sM?$O(B
%%Calculating the integral
%%by the method of steepest descent,
%Combining the contribution from 
%the path along $\varpi$ with that from $e^{2\pi i}\varpi$,
%$\alpha(\omega,\varpi)$
%with that from $\alpha (e^{\pi i}\omega,e^{2\pi i}\varpi)$, 
%\subsubsection{The contribution 
%from the part of $e^{\pi i} \varpi $}
%(i)
%%Since the contributions to $G^C$ 
%%converge in the limit $|\omega| \to m$,
%unlike that from the path along $\varpi $ and $e^{2\pi i}\varpi $.
%Therefore 
%%the contribution
%%is $O(t^{-1})$ at most,
Taking the decay rate  into account,
%at asymptotic late times
we can neglect the contribution from (\ref{t-1tail}) to $G^C$ 
in comparison with
that from (\ref{saddle-int}).
Therefore  tails such as (\ref{saddle-int}) dominate
at asymptotic late times.
%late-time tails can be approximated by
%(\ref{eq:massivetail}).
Finally we arrive at the asymptotic late-time tail as 
\begin{equation}
  \label{eq:massivetail}
 G^{C}(r_{\ast},r'_{\ast};t) 
\simeq  \frac{1}{2\sqrt 3}(2\pi)^{\frac 56}
(mM)^{\frac 13}(mt)^{-\frac 56} 
\sin(mt)
\tilde{\psi}_1(r_{\ast},m)\tilde{\psi}_1(r_{\ast}',m),
%+ \varphi _0)
%\tilde{\psi}_1(r_{\ast},m)\tilde{\psi}_1(r'_{\ast},m),
\end{equation}
%It is  found that 
of which 
the decay rate is independent of the location $r_{\ast}$.

It is  found that 
%the contribution from the saddle point $\omega _0$ dominates,
the time,
when the conditions 
for the application of the saddle-point integration, namely,
%existence of the saddle point (\ref{saddlepoint}), namely,
(\ref{eq:late-time}), (\ref{saddle1}) and (\ref{saddle2}) 
are all satisfied,
%and that the time 
must come sooner or later independent of field mass,
and the tail (\ref{eq:massivetail})
is the asymptotic behavior at late times in the limit $mt \to \infty$.
%In addition, because the time dependence of $\varpi _0$ is common,
%the power of decay caused by the existence of
%non-cancellation phase
%is independent of field mass

%We obtain the contributions to the integral from the parts 
%\begin{equation}
%  \label{eq:massivetail}
% G^{\scriptscriptstyle C} 
%= t^{-1}e^{-imt}f(mM,l)
%\end{equation}
%where $f$ is nontrivial function satisfying $|f|\sim 1$
%But we find that the contributions can be neglected, compared
%with originated from steepest descent. 

%Now we consider which of the tail dominate. in each time scale.
%We find the stage when $t^{-\frac 56 }$ tails dominate 
%appear after the stage when $t^{-\frac 56 }$  tails dominate.
%We find decay rate decelerate step by step.
%$B$Y$-$N;X?t(B$n$$B$O!";~4V$H$H$b$KJQ2=$9$k!#(B
%$B$=$N%?%$%`%9%1!<%k$O(Bmass $B$K0MB8$9$k!#(B

%\qquad\quad  power $n$ varies according to time scale\\
%Relations between field mass and time scale when each tail dominates 
%Therefore we obtain asymptotic late-time tails.
%The steepest descent can be applied if the following conditions;
%\begin{eqnarray*}
%  \label{eq:condition}
%mt &\gg& 1\\
%mt &\gg &\frac{1}{m^2M^2}\\
%mt &\gg& mM  
%\end{eqnarray*}
%are all satisfied. So 

%%%%%%%%%%%%%%%%%%%%%%%%%%%%% Summary %%%%%%%%%%%%%%%%%%%%%%%%%%%%%%%%%%
\section{Summary and discussions}
\label{sec:Summary}
%%%%%%%%%%%%%%%%5 $BF@$i$l$?0lHLE*@-<A$N$^$H$a(B %%%%%%%%%%%%%%%%5

In this paper we have investigated mass-induced behaviors which 
appear in late-time 
tails of classical massive scalar fields
in nearly extreme Reissner-Nordstr\"{o}m background.
If the field mass is small, namely $mM\ll 1$,
the intermediate tails given by Eq. (\ref{eq:HP}) have been shown to dominate 
at the intermediate late-time 
$mM\ll mt \ll 1/(mM)^2$, consistently with 
% the results in
\cite{HandP} (see also \cite{Burko}).
Our main result is the asymptotic tail with 
the decay rate of  $t^{-5/6}$, which is interestingly  independent of
the field mass $m$ and the angular momentum parameter $l$.
This behaviors of inverse power-law decay
supports the numerical results in \cite{Burko}.

%%%%%%%%%%%%%%%%% $B9M;!!'6&LD$K$D$$$F(B $BDI2C(B %%%%%%%%%%%%%%%%%%%%%%%%%

Late-time tail behaviors are generally caused by
the domination of the backscattering from far regions.
It is found the asymptotic tail of massive scalar fields 
(\ref{eq:massivetail}) appears
when the effective contribution to the integral (\ref{eq:branch-cut2})
arises from the region (\ref{rela:asymptotic}),
namely, when the backscattering due to spacetime curvature dominates.
%Therefore, the tail is independent of the parameter $l$.
Further, the frequencies of waves which contribute to  the backscattering
are sharply peaked about 
%are limited to the immediate vicinity of 
%the peculiar value 
$\omega _0$.
%depends on both of field's parameter $m$ and spacetime parameter $M$. 
%are limited to the vicinity of $\omega _0$
%which depends on both of field's parameter $m$ and spacetime parameter $M$.
%dominate
%
These facts suggest that the asymptotic $t^{-5/6}$ tail 
is caused by a {\it resonance} backscattering due to spacetime curvature.
%%%%%%%%%%%%%%%%% $B9M;!!'6&LD$K$D$$$F(B (time scale $B$G8=$l$k$H7kO@(B)%%%%%%%%%%%%%

%We can discuss 
%the relation between the field mass and 
We can also clarify the  resonant picture 
%to discuss 
from a viewpoint of 
the timescale when the $t^{-5/6}$ tail dominates.
%time scale when $t^{-5/6}$ tails begin to dominate.
%We consider 
The basic condition for the tail dominance is given by 
Eq. (\ref{eq:late-time}). %, $mt \gg 1$.
However, for $mM \ll 1$, the $t^{-5/6}$ tail requires 
the additional condition (\ref{rela:asymptotic}), 
which gives $mt \gg 1/m^2M^2$.
%The parameter $mt$ is appropriate for the time scale,
%because it  takes $mt \gg 1$ 
%for tail behaviors to dominate basically.
%All the conditions 
%(\ref{eq:late-time}), (\ref{saddle1}) and (\ref{saddle2})
%should be for the tail behavior.
%Then for small field mass, $mM \ll 1$,
%the smaller $mM$ is,
%the later the tail begin to dominate as $mt \gg 1/m^2M^2$.
On the other hand, 
for large  field mass $mM \gg 1$,
the larger $mM$ is,
the later the $t^{-5/6}$ tail begins to dominate,
because the tail requires the further condition
$\varpi _0/m \ll 1$ for Eq. (\ref{saddlepoint})
in addition to the conditions 
(\ref{eq:late-time}) and (\ref{rela:asymptotic}),
which gives $mt \gg mM$.
Therefore the timescale
when the $t^{-5/6}$ tail dominates
will become minimum
%soonest 
at  $mM \simeq O(1)$,
%In other words, 
which means that the most effective backscattering occurs 
for such massive scalar fields at late times.
%which seems to be 
%caused by that resonance between field mass and space-time curvature
%occurs most efficiently.
%This means scalar fields 
%remain the most, if $mM \simeq 1$. 
Here we note that 
%noting that 
the timescale when the $t^{-5/6}$ tail dominates
is determined by the black-hole radius $M$.
It is conjectured that 
%the $t^{-5/6}$ tail 
the slow decay of $t^{-5/6}$
%of scalar fields 
originates from %effective slow decay of scalar fields by 
the existence of resonant enhancement of massive scalar waves 
with the peculiar frequency $\omega _0$ %$\omega \simeq m$ 
near the horizon. 
%In this sense
In conclusion,
we claim that 
resonance behaviors in field-mass dependence are not 
peculiar to quantum scalar fields but manifest also
in classical scalar fields.

%%%%%%%%%%%%%%% Burko and Ori $B$K$D$$$F(B %%%%%%%%%%%

%The determination of the correct late-time power-law indeces
%is essential for the investigation the black hole's interor.
%Generally, it is difficult for 
%numerical methods to evaluate the evolution accuracy
%at very late times.

%In \cite{B-O},
%they have made 
%to determine the power-law indices in the nonlinear 
%collapse problem.
%Their scheme is convenience
%to study the evolution to asymptotically late times.

%%%%%%%%%%%%%%% $B2]Bj(B %%%%%%%%%%%%%%%

In this paper we have calculated the tail behaviors in nearly
extremal limit,
because we are motivated by the  previous work \cite{TandK}.
The extension of our calculation into non-extremal case remains in future works,
to investigate 
whether 
our result in this paper is a special feature of the extremal case or not.
\\

\acknowledgments

The authors thank A. Hosoya, M. Sakagami, A. Ohashi and  
colleagues of CG laboratory in Nagoya University
for valuable discussions and comments.
H.K. thanks T. Konishi and Y. Hirata for useful advice 
on analysis of differential equations
%on the earlier stage of the calculation 
and 
%HK also 
also thanks 
A. Yamada for 
drawing the figure in this paper.

\end{document}